\documentclass{article}
\usepackage{spconf,amsmath,graphicx}
\usepackage{multirow}
\usepackage{url}
\usepackage{amssymb}
\usepackage{comment}
\usepackage{balance}
\usepackage{fancyhdr}

\fancypagestyle{FirstPage}{
\lfoot{\scriptsize{© 2023 IEEE.  Personal use of this material is permitted.  Permission from IEEE must be obtained for all other uses, in any current or future media, including reprinting/republishing this material for advertising or promotional purposes, creating new collective works, for resale or redistribution to servers or lists, or reuse of any copyrighted component of this work in other works.}} 
}

\let\OLDthebibliography\thebibliography
\renewcommand\thebibliography[1]{
  \OLDthebibliography{#1}
  \setlength{\parskip}{0pt plus 3pt}
  \setlength{\itemsep}{0.3pt plus 3pt}
}

\ninept 

\title{Adversarial Guitar amplifier modelling with unpaired data}
\name{Alec Wright$^1$, Vesa V\"{a}lim\"{a}ki\sthanks{\scriptsize This research is part of the activities of the Nordic Sound and Music Computing Network---NordicSMC (NordForsk project no.~86892)}$^{,1}$, Lauri Juvela$^{1,2}$}
\address{$^1$Acoustics Lab, Dept.~Information \& Communications Eng.	Aalto University, Espoo, Finland \\ $^2$Neural DSP Technologies,
 Helsinki, Finland}
\begin{document}
%

\maketitle

\begin{abstract}
We propose an audio effects processing framework that learns to emulate a target electric guitar tone from a recording. We train a deep neural network using an adversarial approach, with the goal of transforming the timbre of a guitar, into the timbre of another guitar after audio effects processing has been applied, for example, by a guitar amplifier. The model training requires no paired data, and the resulting model emulates the target timbre well whilst being capable of real-time processing on a modern personal computer. To verify our approach we present two experiments, one which carries out unpaired training using paired data, allowing us to monitor training via objective metrics, and another that uses fully unpaired data, corresponding to a realistic scenario where a user wants to emulate a guitar timbre only using audio data from a recording. Our listening test results confirm that the models are perceptually convincing.

\end{abstract}
\begin{keywords}Audio systems, deep learning, generative adversarial networks, music, nonlinear systems, unsupervised learning.
\end{keywords}
\section{Introduction}
\thispagestyle{FirstPage}
\label{sec:intro}

When recording musical instruments for music production, audio effects are an essential component of the resulting timbre. This is especially true of electric guitar, where the timbre, or as it is often referred to by guitarists, \textit{tone}, imparted by certain amplifiers and effects pedals is highly sought-after. Digital emulation of analog audio devices such as these is known as Virtual Analog (VA) modelling \cite{Valimaki11}. 

A related but more challenging problem is to emulate the timbre of an electric guitar directly from a recording. This corresponds to a practical and very exciting task: how to imitate the guitar tone heard on a commercial music recording using your own electric guitar? We refer to this as modelling from \textit{unpaired data}, as the unprocessed guitar signal used to generate the target guitar tone is unavailable.

The usual paradigm of VA modelling is the emulation of certain analog devices, however this is insufficient for this purpose for two reasons. 
Firstly, the specific instrument and devices used to craft the guitar tone on a famous recording might not be known. Even if the required equipment is available, recreating the recording setup is a time consuming process that requires expert knowledge.

Secondly, the desired timbre includes both the musical instrument itself, as well as any processing applied to it. This means that any emulation of a recorded guitar tone must also account for differences between the timbre of your guitar, and the timbre of the target guitar, before effects processing is applied.

The problem thus corresponds more closely to that of timbre transfer, but in the specific case where the input and target instrument are both guitars. The problem can be  divided into two sub-problems: the guitar-to-guitar transformation, and modelling of the guitar signal chain. Previous work has approached guitar-to-guitar transformation using linear filtering, for example, to process an electric guitar such that it sounds like an acoustic guitar \cite{Karjalainen00a}.


The guitar signal chain modelling problem involves the emulation of all the devices, other than the instrument itself, used to create the recording. These might include a guitar amplifier, speaker cabinet, and other effects such as compression or distortion. Modelling these types of devices has been widely studied, and broadly falls into two approaches, ``white-box'', where equations describing the system behaviour are derived
\cite{Karjalainen06, YehIEEETrans, kroning2011analysis}, or ``black-box'', where data collected from the system is used to fit a generic model \cite{novak2010chebyshev, eichas2016black, orcioni2018identification}. Recently, deep learning methods based on neural networks have become a popular choice \cite{vanhatalo2022review}. 
These include architectures for modelling a wide range of effects \cite{ramirez2019modeling}, feedforward \cite{damskaggICASSP}, and recurrent \cite{WrightRNN19} models for guitar amplifiers and time-varying effects \cite{wright2021neural}, as well as models using DSP components such as linear filters \cite{kuznetsov2020differentiable, nercessian2021lightweight, wright2022grey}.

Whilst the aforementioned approaches are all viable for the task of imitating a target electric guitar timbre, they all require us either to know what devices were used during recording, or to have access to \textit{paired data} with both the target timbre and the unprocessed signal taken directly from the guitar used during the recording. In the case where paired training data is impossible, or unavailable, one solution is to formulate the problem as domain transfer, similar to \textit{Image Style Transfer}, where the style of an input image is transformed to match a target style, whilst retaining the input image content \cite{NeuralStyleTransRev, cycleGAN}.  

A related problem in audio is \textit{Audio Style Transfer}, which seeks to transfer the style of some input audio to a target style, whilst retaining the content of the input audio \cite{AudStyTrans}. In the speech domain, an example of this is voice conversion, 
which can be achieved, for example, using non-parallel speech data to train a Generative Adversarial Network (GAN) with a cycle-consistency loss \cite{Kaneko2018-cycle-gan-voice-conversion}.


For musical applications the problem is frequently referred to as \textit{timbre transfer} \cite{huang2018timbretron, jain2020att} or \textit{tone transfer} \cite{carney2021tone}. 
This can be achieved by applying image style transfer techniques to a time-frequency representation of the input audio, and then re-synthesising the audio in the waveform domain \cite{huang2018timbretron}, or by learning a high level representation of the input and using that as input to a synthesizer \cite{DDSP}. Recent work has also proposed style transfer of audio effects, to impart the production style of one recording onto another \cite{steinmetz2022style}.


\begin{figure}[]
    \includegraphics[width=\columnwidth]{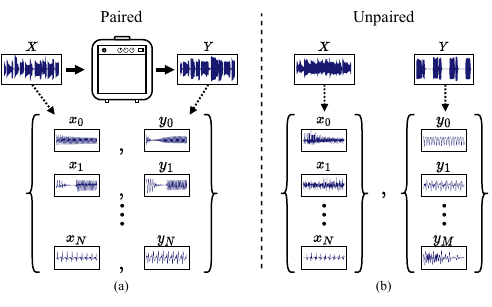}
        \vspace{-25pt}
    \caption{(a) Supervised black-box modelling is based on \textit{paired} audio data $\{x_i, y_i\}^N_{i=0}$, where the target audio $y_i$ is obtained by processing the input audio $x_i$ with the target device. When paired data is unavailable, we propose to use (b) \textit{unpaired} data, made up of examples of a \textit{source timbre} $\{x_i\}^N_{i=0} \in X$  and examples of a \textit{target timbre} $\{y_j\}^M_{j=0} \in Y$, where neither the content nor the timbre contained in $x_i$ match those contained in $y_j$.}
    \label{fig:paireddata}
        \vspace{-10pt}
\end{figure}


In this paper, we present a method to process a signal recorded directly from an electric guitar, or \textit{input timbre}, such that its timbre matches that of a guitar from a recording, or \textit{target timbre}. We solve this domain transfer problem using an unsupervised approach, in which a black-box feedforward convolutional model is trained adversarially using unpaired data. The distinction between supervised and unsupervised modelling is shown in Fig.~\ref{fig:paireddata}. 



The rest of this paper is structured as follows. Sec.~\ref{sec:method} discusses the modelling approach, training objective and data. Sec.~\ref{sec:evaluation} describes two experiments conducted to validate the proposed approach. Sec.~\ref{sec:res} discusses the results and Sec.~\ref{sec:conclusion} concludes the paper.

\section{Method}
\label{sec:method}

The proposed method uses an adversarial training approach. We define domains in the space of guitar timbre, where a domain is defined by the sounds produced by a particular combination of a guitar, pickups, amplifier, and any other audio effects. Note that the domain is defined purely by the guitar timbre, and not the content of the guitar playing. If we define domains, $X$ and $Y$, and we sample audio data, $x$, where $x \in X$, and $y$, where $y \in Y$, our objective is to learn a mapping, $G$, that converts audio from domain $X$ to domain $Y$:
\begin{equation}
	G(x) = \hat{y}.
	\label{Mapping}
\end{equation}
\noindent The ultimate objective is to transform the tone such that $\hat{y}$ is perceptually indistinguishable from $y$. To train $G$ we use a discriminator model $D$ trained to identify examples from the target timbre domain.

Previous work on adversarial domain-to-domain translation \cite{cycleGAN, Kaneko2018-cycle-gan-voice-conversion} has applied a cycle-consistency criterion to avoid a specific form of mode collapse,
which results in $G$ simply ignoring the content of the input $x$ and synthesising unrelated content from the target domain.
However, in our present experiments the generator model $G$ did not exhibit this problem, likely due to the constrained expressive capability of the feedforward convolutional neural network used, which learns to directly apply a transformation to the input signal in the time domain.

\subsection{Generator}

The generator model $G$ used in this work is a feedforward variant of the WaveNet architecture \cite{oord2016wavenet}. A non-causal feedforward WaveNet variant was first proposed for speech denoising \cite{rethage2018wavenet}. A causal version was later applied to guitar amplifier modelling \cite{damskaggICASSP}, and this architecture is used as the generator throughout this work. The model has two main components, a stack of dilated causal convolutional layers, and a linear post-processor. The post-processor is a fully connected layer that takes the outputs of each of the convolutional layers as input. 

All generator models used in this work consist of two stacks of nine dilated convolutions, with each stack starting at a dilation of one and increasing by a factor of two with each subsequent convolutional layer. Each convolutional layer has a kernel size of three, and uses the same gated activation function as the original WaveNet \cite{oord2016wavenet}. The receptive field of this model is 2045 samples, or about 46.4 ms at a 44.1-kHz sample rate. It should also be noted that previous work has shown that a C++ implementation of this model is capable of running in real-time on a modern desktop computer \cite{damskagg2019real}.

\subsection{Discriminator}

The input to the proposed discriminator $D$ is a time-frequency representation of the audio, as proposed in \cite{juvela2018-synthesis-from-mfcc}. The discriminator consists of a stack of 1D convolutional layers, and the frequency bins of the time-frequency representation are provided as channels to the first layer of the discriminator. Subsequent layers use grouped convolutions to reduce computational cost. The hyperparameters for each layer are shown in Table \ref{table:DiscParameters}. All layers use weight normalization, and all layers except the final output layer are followed by a Leaky ReLU activation function with negative slope of 0.2. Four different time-frequency representations were trialled, either a magnitude spectrogram, a magnitude mel-spectrogram, a log magnitude spectrogram, or a log magnitude mel-spectrogram. For all mel-spectrograms, 160 mel bands were used, the maximum frequency was set to Nyquist and the minimum frequency was set to 0\,Hz.

Additionally, a multi-scale version of the spectral domain discriminator was trialled which included three sub-discriminators, each operating on time-frequency representations obtained using different window sizes. In the case of the single spectral-domain discriminator, a window size of $N = 1024$ was used, as for the multi-scale spectral discriminator, window sizes of $[512, 1024, 2048]$ were used. In all cases, the hop size was set to $N/4$.

\begin{table}[t!]
\caption{Convolutional layer parameters for proposed spectral domain discriminator.}
\vspace{6pt}
\resizebox{\columnwidth}{!}{
\begin{tabular}{l|rrrrrrr}
\hline
Layer \#                          & 1  & 2   & 3   & 4    & 5    & 6    & 7 \\ \cline{2-8} \hline
\multicolumn{1}{l|}{Kernel Size}  & 10 & 21  & 21  & 21   & 21   & 5    & 3 \\
\multicolumn{1}{l|}{Out Channels} & 32 & 128 & 512 & 1024 & 1024 & 1024 & 1 \\
\multicolumn{1}{l|}{Groups}       & 1  & 8   & 32  & 64   & 64   & 1    & 1 \\
\hline
\end{tabular}}
\vspace{-12pt}
\label{table:DiscParameters}
\end{table}

\subsection{Training Objective}

\begin{figure}[t!]

    \includegraphics[width=\columnwidth]{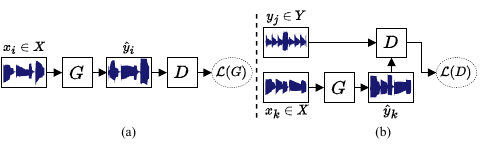}
    \vspace{-22pt}
    \caption{Training setup for (a) the Generator and (b) Discriminator. Generator inputs, taken from the input domain $X$, are processed to emulate the timbre, but not content, of the target domain $Y$.}
    \label{fig:train}
        \vspace{-15pt}
\end{figure}

The generator and discriminator of the GAN were trained using the hinge loss objective \cite{lim2017geometric}, identical to that which was used in MelGAN \cite{kumar2019melgan}, as follows:
\begin{equation}
    \mathcal{L}(D) = \mathbb{E}_y[\textrm{max}(0, 1 - D(y))] + \mathbb{E}_x[\textrm{max}(0, 1 + D(G(x)))]
    \label{adv_loss_disc}
\end{equation}
\noindent and
\begin{equation}
    \mathcal{L}(G) = \mathbb{E}_x[- D(G(x))],
    \label{adv_loss_gen}
\end{equation}

\noindent respectively, where $x$ is a guitar audio waveform used as input to the generator and $y$ is a guitar audio waveform taken from the target set. The training scheme is shown in Fig.~\ref{fig:train}. During training, both the input and target guitar datasets are split into two-second segments before processing by $G$ or $D$. The models were trained using the Adam optimizer \cite{kingma2014adam} with a batch size of 5.

\subsection{Data}

 
The data used throughout this work was taken from a guitar dataset 
originally proposed for the task of automatic transcription \cite{kehling2014automatic}. We use the fourth subset of the dataset, which consists of 64 short musical pieces played at fast and slow tempi. The pieces were recorded on two different electric guitars, a Career SG and an Ibanez 2820. We processed the recordings to remove leading and trailing silence, as well as the 2 bar count in at the beginning of each piece. Additionally, it was noted that clipping is present in some of the samples in the Career SG dataset, so examples where excessive clipping was observed were removed.

After the pre-processing, there was approximately 40 min of audio from the Ibanez 2820 guitar, and 30 min from the Career-SG. To create the datasets used during our experiments, the guitar audio was processed by a guitar amplifier plugin. To test the robustness of our modelling approach, a separate dataset was created for three different plugin settings, hereafter referred to as `Clean', `Light Distortion', and `Heavy Distortion'. The amount of harmonic distortion introduced increases from a relatively small amount in the `Clean' setting, to an extremely distorted guitar tone found in the `Heavy Distortion' case. 

\section{Experiments}
\label{sec:evaluation}

Our proposed modelling approach was tested on two different problems. In the first scenario the signal contained in both the input and target datasets is recorded from the same guitar. In this case, the specific instrument in each dataset is identical, and the modelling task is to recreate the effects processing applied to the input signal. Although not realistic, this scenario is relevant because it allows the use of supervised metrics to evaluate our unsupervised training method.

In the second (more realistic) scenario, the input dataset is the unprocessed audio recorded from one guitar, and the target dataset is audio that has been recorded from a different guitar, after audio effects processing has been applied to it. In this case, the modelling task implicitly includes transforming the tone of one guitar into another, as well as recreating the effects processing applied to that guitar. The two experiments are depicted in Fig.~\ref{fig:datasetup}. 

\subsection{Experiment 1: Single Guitar}

The dataset was split to ensure that the input and target datasets do not contain any of the same guitar content, by dividing it into two-second segments, with subsequent segments being sent alternately to either the input or target training dataset. This ensures that the spectral content of the unprocessed guitar in both datasets is similar, but that the actual content is different. 

Unsupervised models trained with the spectral domain discriminator configurations introduced in Sec. \ref{sec:method} were tested, as well as an unsupervised baseline that used the MelGAN discriminator \cite{kumar2019melgan}. As a supervised baseline, the same generator model was trained in a supervised manner, to minimise the Error-to-Signal ratio loss function, $\mathcal{E}_\text{ESR}$, with high-pass filter pre-emphasis \cite{damskaggICASSP}.

\begin{figure}[t!]
    \centering
    \includegraphics[width=0.95\columnwidth]{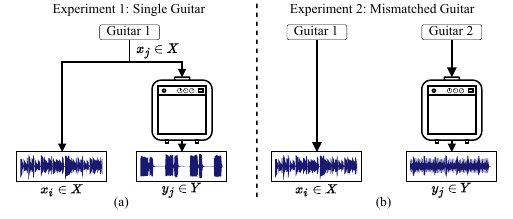}
    \vspace{-0.5cm}
    \caption{(a) In the \textit{Single Guitar} experiment, the input and output audio are produced using the same guitar, but with unpaired data, whereas (b) in the \textit{Mismatched Guitar} experiment, the input and output audio are generated with different guitars.}
    \label{fig:datasetup}
    \vspace{-0.5cm}
\end{figure}

As in this case the guitar is the same in both datasets, as shown in Fig.~\ref{fig:datasetup}(a), a ground truth, or reference, for how the input guitar should sound after the effects processing is applied is available. This allows a validation loss to be calculated over a held-out validation set, that consists of paired input/output guitar audio. 

For validation loss metrics, we use both the linear and log scaled multi-scale magnitude spectrogram loss described in \cite{DDSP}, which we refer to as $\mathcal{E}_\text{ms}$ and $\mathcal{E}_\text{lms}$ respectively. In addition to this, we also present the L1 distance between the output and target mel magnitude spectrograms, again using both linear and log scaling, which we will refer to as $\mathcal{E}_\text{mel}$ and $\mathcal{E}_\text{lmel}$ respectively.

Additionally, a MUSHRA \cite{ITUMUSHRA} style listening test was carried out. Participants were presented with audio clips that were processed by the target plugin, as well as the various neural network models. An anchor was also included, which was created by processing the input with a tanh nonlinearity, as well as a hidden reference. Participants were asked to rate each test condition out of 100, based on perceived similarity to the reference.

Results are shown in Table \ref{table:Exp1Results}. In each case the training was run for 400k iterations, and the validation loss was used to select the best performing model. For the spectral domain discriminators, the validation loss used to determine the best performing model was chosen depending on the form of the input provided to the model, for example, if the input to the discriminator was a log-mel spectrogram, then the training iteration where the lowest validation $\mathcal{E}_\text{lmel}$ was achieved was selected as the best performing model. In each case, our experiments included both multi-scale and single-scale spectral domain discriminators, however, for brevity, only the results for the best performing of the two are included in Table \ref{table:Exp1Results}.

\subsection{Experiment 2: Mismatched Guitar}

For the scenario in Experiment 2, the guitar used to create each dataset is different, as shown in Fig.~\ref{fig:datasetup}(b). This means that objective metrics 
are unavailable. As such, we conducted a listening test, in which participants were presented with a reference, consisting of a few seconds of guitar playing from the target timbre domain. The participants were then asked to rate a number of test conditions, which consisted of the next few seconds of the same piece of music, but performed on the guitar from the input timbre domain. The test conditions all consisted of processed versions of the same guitar audio. It was impossible to include a hidden reference in the test, as it does not exist.

Two baselines were created, both having access to some ground truth information. The first baseline was created by processing the input guitar with the same effects plugin that was used to create the target guitar timbre, this baseline is referred to as the ``plugin-only'' timbre. This corresponds to a simplified solution in which the guitar-to-guitar transformation is not included, but the signal chain applied to the guitar is identical to that used to produce the reference. 

The second baseline, referred to as the ``EQ+plugin'', was created by applying a linear equalization (EQ) matching to the unprocessed input guitar tone, with the EQ target being the target guitar before effects processing was applied. This EQ-matched version of the input guitar was then processed by the effects plugin used to create the reference timbre. 
Notice that this processing is impossible to achieve in a practical setting, but is used here in lieu of an ideal reference. A low-quality anchor was also included in the listening test, which consisted of the input guitar processed by a tanh nonlinearity.

The test conditions consisted of three unsupervised models, selected based on the results of the first experiment. These were trained using the MelGAN discriminator, or the spectral domain discriminator with either log spectrogram or log mel-spectogram input. In both cases, the multi-scale version of the spectral domain discriminator was used. The models were trained for 250k iterations. The results of the listening test are shown in Fig.~\ref{fig:mushra}.

The listening tests for both experiments were completed by twelve participants. Three participants identified as female and nine as male. The participants all had experience completing listening tests, and their mean age was 30.8 years. For the first listening test, four participants were removed in post screening as they rated the hidden reference less than 90 in more than $15\%$ of the trials.

\section{Discussion}
\label{sec:res}

\begin{table}[t!]
\vspace{-9pt}
\caption{Objective and subjective results for the Single Guitar experiment. For validations losses, bold indicates best performing unsupervised model. For the listening test result bold indicates best performing of all models and $95\%$ confidence intervals are shown.}
\label{table:Exp1Results}
\vspace{3pt}
\resizebox{\columnwidth}{!}{
\begin{tabular}{cccccccc}
\hline
\multicolumn{1}{l}{} & \multicolumn{1}{l}{}          & \multicolumn{5}{c}{Validation Loss}                                                                                                            & Listening            \\
\multicolumn{2}{c}{Model}                            & $\mathcal{E}_{\text{ms}}$ & $\mathcal{E}_{\text{lms}}$ & $\mathcal{E}_{\text{mel}}$ & $\mathcal{E}_{\text{lmel}}$ & $\mathcal{E}_{\text{ESR}}$ & Test                 \\ 
\hline \hline
\multicolumn{8}{c}{Target Tone: Clean}                                                                                                                                                                                       \\ \hline
\multicolumn{2}{c|}{Supervised}                      & 5.12                      & 0.76                       & 0.57                       & 0.12                        & \multicolumn{1}{c|}{0.003} & 81$\pm$4.1          \\
\multicolumn{2}{c|}{MelGAN}                          & \textbf{37.5}                      & 1.47                       & \textbf{2.75}              & \textbf{0.17}               & \multicolumn{1}{c|}{2.38}  & 71$\pm$4.8           \\ \hline
\multicolumn{2}{c|}{Spectral Domain}                 & \multicolumn{1}{l}{}      & \multicolumn{1}{l}{}       & \multicolumn{1}{l}{}       & \multicolumn{1}{l}{}        & \multicolumn{1}{l}{}       & \multicolumn{1}{l}{} \\
Input                & \multicolumn{1}{c|}{\# Disc.} & \multicolumn{1}{l}{}      & \multicolumn{1}{l}{}       & \multicolumn{1}{l}{}       & \multicolumn{1}{l}{}        & \multicolumn{1}{l}{}       & \multicolumn{1}{l}{} \\ \hline
Spect.               & \multicolumn{1}{c|}{1}        & 39.2                      & 3.27                       & 3.39                       & 0.39                        & \multicolumn{1}{c|}{2.55}  & 32$\pm$4.7           \\
Mel                  & \multicolumn{1}{c|}{1}        & 40.0                      & 1.51                       & 2.88                       & 0.28                        & \multicolumn{1}{c|}{1.27}  & 46$\pm$4.4           \\
Log Spect.           & \multicolumn{1}{c|}{3}        & 44.1                      & \textbf{0.81}                       & 3.76                       & 0.18                        & \multicolumn{1}{c|}{2.71}  & 82$\pm$4.5           \\
Log Mel              & \multicolumn{1}{c|}{3}        & 46.9                      & 0.93                       & 4.07                       & 0.19                        & \multicolumn{1}{c|}{\textbf{1.04}}  & \textbf{83$\pm$3.9}           \\
\hline \hline
\multicolumn{8}{c}{Target Tone: Light Distortion}                                                                                                                                                                            \\ \hline
\multicolumn{2}{c|}{Supervised}                      & 2.57                      & 0.81                       & 0.28                       & 0.09                        & \multicolumn{1}{c|}{0.001} & \textbf{93$\pm$3.0}           \\
\multicolumn{2}{c|}{MelGAN}                          & \textbf{25.2}                      & 2.18                       & \textbf{1.32}                       & \textbf{0.18}                        & \multicolumn{1}{c|}{2.51}  & 73$\pm$5.4           \\ \hline
\multicolumn{2}{c|}{Spectral Domain}                 &                           &                            &                            &                             &                            &                      \\
Input                & \multicolumn{1}{c|}{\# Disc.} &                           &                            &                            &                             &                            &                      \\ \hline
Spect.               & \multicolumn{1}{c|}{1}        & 32.5                      & 4.26                       & 2.39                       & 0.45                        & \multicolumn{1}{c|}{\textbf{1.49}}  & 35$\pm$4.0           \\
Mel                  & \multicolumn{1}{c|}{1}        & 34.4                      & 4.12                       & 2.57                       & 0.48                        & \multicolumn{1}{c|}{2.43}  & 34$\pm$4.0           \\
Log Spect.           & \multicolumn{1}{c|}{1}        & 45.3                      & \textbf{1.11}                       & 4.51                       & 0.23                        & \multicolumn{1}{c|}{2.18}  & 81$\pm$4.8           \\
Log Mel              & \multicolumn{1}{c|}{3}        & 38.1                      & 1.17                       & 3.36                       & 0.21                        & \multicolumn{1}{c|}{2.50}  & 88.7$\pm$3.9         \\
\hline \hline
\multicolumn{8}{c}{Target Tone: Heavy Distortion}                                                                                                                                                                            \\ \hline
\multicolumn{2}{c|}{Supervised}                      & 6.33                      & 2.53                       & 0.60                       & 0.19                        & \multicolumn{1}{c|}{0.03}  & 57$\pm$4.6           \\
\multicolumn{2}{c|}{MelGAN}                          & \textbf{22.4}                      & \textbf{2.49}                       & \textbf{1.81}                       & \textbf{0.22}                        & \multicolumn{1}{c|}{2.04}  & \textbf{92$\pm$2.8}           \\ \hline
\multicolumn{2}{c|}{Spectral Domain}                 &                           &                            &                            &                             &                            &                      \\
Input                & \multicolumn{1}{c|}{\# Disc.} &                           &                            &                            &                             &                            &                      \\ \hline
Spect.               & \multicolumn{1}{c|}{1}        & 28.9                      & 4.14                       & 2.70                       & 0.37                        & \multicolumn{1}{c|}{2.33}  & 54$\pm$5.7           \\
Mel                  & \multicolumn{1}{c|}{1}        & 25.5                      & 7.15                       & 2.36                       & 0.60                        & \multicolumn{1}{c|}{\textbf{0.86}}  & 28$\pm$3.4           \\
Log Spect.           & \multicolumn{1}{c|}{1}        & 32.1                      & 2.52                       & 3.25                       & 0.29                        & \multicolumn{1}{c|}{3.17}  & 81$\pm$4.8           \\
Log Mel              & \multicolumn{1}{c|}{3}        & 24.5                      & 2.55                       & 2.21                       & 0.23                        & \multicolumn{1}{c|}{2.37}  & 85$\pm$3.8           \\
\hline
\end{tabular}
} 
\end{table}

The objective results for Experiment 1, shown in Table \ref{table:Exp1Results}, indicate that the feedforward WaveNet model trained in a supervised fashion performs better than the unsupervised models on all the proposed metrics. Generally, of the unsupervised models, those trained with the MelGAN discriminator tend to perform better in the objective loss metrics. However, the results from the listening tests for Experiment 1, also shown in Table \ref{table:Exp1Results}, indicate that there is no clear best performing model between the supervised and unsupervised training approach. The results do clearly show, however, that models trained using a spectral domain discriminator with linear scaled spectrogram as input perform poorly. The supervised model of the `Heavy Distortion' tone also reveals a misalignment between the time-domain ESR loss metric and perception.  For all target tones, at least one unsupervised model achieved a score of 80 or higher, indicating a perceptual match between Good and Excellent on the MUSHRA scale.

\begin{figure}[t!]
    \includegraphics[width=\columnwidth]{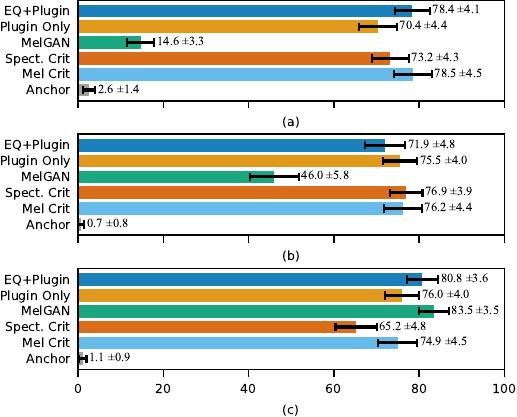}
    \vspace{-19pt}
    \caption{MUSHRA scores with 95\% confidence intervals for the (a) Clean, (b) Light Distortion and (c) Heavy Distortion guitar tone settings that were modelled in Experiment 2.}
    \label{fig:mushra}
    \vspace{-15pt}
\end{figure}

The results of the listening test for Experiment 2, shown in Fig.~\ref{fig:mushra}, indicate that the unsupervised models are competitive with our proposed baselines. For the `Clean' and `Light Distortion' case, the MelGAN model performs poorly. One possible explanation for this is that during training for the first experiment it was observed that the MelGAN produced some oscillation and instability as training went on, as the spectral discriminator models tended to quickly plateau and then remain stable. As no validation loss was available to monitor the training for the mismatched guitar case, it was not possible to select the best performing model after training.

The experimental results show that the proposed framework is capable of producing perceptually convincing models of nonlinear effects processing using unpaired data. The limitations of the proposed method are that for each desired input target pair, a new generator model must be trained. The amount of training time required for a model to achieve a perceptually convincing result was not investigated extensively, but typically took between 4 and 10 hours of training on a GPU. Further work is also required to test the method when applied to real-world use cases, for example when a smaller amount of data from the target guitar timbre is available.

\section{Conclusion}
\label{sec:conclusion}

This work shows for the first time how the guitar timbre heard on a music recording can be imitated by another guitar, using an unsupervised method based on a GAN framework. We formulated the problem as domain transfer, and proposed a spectral domain discriminator. We validated our method through two listening tests and showed that the models produced are perceptually convincing. Audio samples are available at our demonstration page\footnote{\tiny\url{https://ljuvela.github.io/adversarial-amp-modeling-demo}}.


\clearpage
\balance
\setlength{\itemsep}{0pt}
\bibliographystyle{IEEEbib}
\bibliography{strings,refs}

\end{document}